\documentclass[%
 aip,
amsmath,amssymb,
 reprint,%
]{revtex4-1}

\usepackage{graphicx}
\usepackage{dcolumn}
\usepackage{bm}

\usepackage[utf8]{inputenc}
\usepackage[T1]{fontenc}
\usepackage{mathptmx}
\usepackage{etoolbox}
\usepackage{color}

\makeatletter
\def\@email#1#2{%
 \endgroup
 \patchcmd{\titleblock@produce}
  {\frontmatter@RRAPformat}
  {\frontmatter@RRAPformat{\produce@RRAP{*#1\href{mailto:#2}{#2}}}\frontmatter@RRAPformat}
  {}{}
}%
\makeatother
\begin{document}

\preprint{AIP/123-QED}

\title{Thouless pumping of solitons in a nonlocal medium}

\author{Fangwei Ye}
\affiliation{ 
School of Physics, Chengdu University of Technology, Chengdu 610059, China
}%
 \affiliation{School of Physics and Astronomy, Shanghai Jiao Tong University, Shanghai 200240, China}
\email{fangweiye@sjtu.edu.cn}
\author{Aidar H. Ryazhapov}%
\affiliation{Institute of Spectroscopy, Russian Academy of Sciences, 108840 Troitsk, Moscow, Russia}
\affiliation{Faculty of Physics, Higher School of Economics, 105066 Moscow, Russia}
\author{Yaroslav V. Kartashov}
\affiliation{Institute of Spectroscopy, Russian Academy of Sciences, 108840 Troitsk, Moscow, Russia}
\author{Vladimir V. Konotop}
\affiliation{Departamento de Física and Centro de Física Teórica e Computacional, Faculdade de Ciências, Universidade de Lisboa, Campo Grande, Edifício C8, Lisboa 1749-016, Portuga}
\date{\today}

\begin{abstract}

Thouless pumping is a fundamental phenomenon recognized as being widespread across various areas of physics, with optics holding a particularly prominent role. 
Here, we study this effect for optical solitons in a medium where the refractive index is shaped by two slowly sliding sublattices and a nonlocal nonlinear response. The spectral bands of such a potential can exhibit nontrivial topology, and excitations occupying these bands can undergo quantized transport governed by the space -- time Chern indices of the linear spectrum. We find that nonlocality of the medium profoundly affects the dynamics of Thouless pumping. Thus, we show that broad, low-power fundamental solitons do not exhibit transport, as they excite only a small portion of the spectral band, while high-power solitons with broader spectral projections do demonstrate stable quantized transport. The transition point between these two principally different light propagation regimes strongly depends on the degree of nonlocality of the nonlinear response and shifts to larger powers with increasing nonlocality. 
Notably, even a moderate level of nonlocality is sufficient to prevent the breakdown of topological transport at high powers commonly observed in local Kerr media. Beyond fundamental solitons, we also demonstrate that multipole solitons, such as dipole and tripole states can be pumped stably. This is the first time such complex soliton states have been shown to undergo Thouless pumping. While fundamental solitons require only exceeding a power threshold, multipoles exhibit stable transport only within a finite power window. This window is broader for dipoles than for tripoles and expands with increasing nonlocality, revealing a trade-off between structural complexity and stability.

\end{abstract}

\maketitle

\section{\label{sec:level1}Introduction }

Optical systems represent a universal and versatile testbed for observing and studying topological phenomena, particularly enabling their detailed exploration in the presence of nonlinearity. The quantized transport, predicted by Thouless~\cite{Thouless1983} for electrons and recognized as a ubiquitous wave phenomenon observed across many physical disciplines~\cite{Citro2023}, is one of them. This phenomenon is attracting significant attention nowadays. Topological pumping of light in linear systems has been observed in waveguide arrays~\cite{Verbin2015, Zilberberg2018, Cerjan2020, Jurgensen2021, Sun2022, Jurgensen2023}, optical lattices induced in photorefractive crystals~\cite{Wang2022,Yanga2024}, and in resonator arrays~\cite{Tangpanitanon2016}. The possibility to observe pumping in optical settings naturally raises questions about the impact of nonlinearity on the pumping process, particularly because nonlinearity of a medium can lead to the formation of solitons. Pumping in soliton bearing systems has been addressed in a number of recent studies. The cubic (alias Kerr, {\color{black} local}) nonlinearity in the governing nonlinear Schr\"odinger (NLS) equation is the most studied one. It is known that it may result in breaking of topological transport, as well as in fractional transport~\cite{Jurgensen2021, Fu2022a, Jurgensen2023} in one-dimensional settings. In two-dimensional settings, Thouless pumping can be accompanied by splitting of the initial state into multiple solitons exhibiting complex dynamical scenarios~\cite{Fu2022b}. 
{\color{black}
In a continuous medium with a band-gap spectrum some of the phenomena mentioned above can be explained by the exchange of energy between different bands during the pumping process — a mechanism that is especially relevant for high-intensity pulses. In discrete systems, observing such phenomena requires constructing arrays with complex unit cells containing different waveguides, thereby characterized by multiple spectral bands. So far, studies of local nonlinearities have shown that they can indeed couple different bands.  
}
 
The diversity of solitons is not restricted to one-component systems, that stimulated studies of pumping of two-component solitons in coupled NLS equations in discrete \cite{Mostaan2022} and continuous~\cite{Lyu2024} lattices as well as in quadratic nonlinear media~\cite{Kartashov2025}, where fields in different components can propagate in effectively different media. {\color{black} For an overview of the nonlinear Thouless pumping in discrete optical systems see the recent review~\cite{Szameit2024}.}

In this paper, we report on Thouless pumping of solitons in a nonlocal medium, a scenario significantly different from those considered in previous studies, resulting in distinct dynamics.  {\color{black} Such a nonlinearity introduces an additional spatial scale into the problem: the width of the nonlocal kernel, which characterizes the spatial extent of the medium “felt” by the soliton. This scale must be compared to the natural length scale introduced by the lattice periodicity. Furthermore, in nonlocal media, the dependence of the soliton amplitude on the total power differs significantly from that in the local case, which is expected to suppress inter‑band transitions even at high powers. This can have a profound impact on the observability of phenomena such as fractional pumping or its breakdown. Moreover, the stability of the solitons may be strongly affected.}  Specifically, we demonstrate that pumping of fundamental solitons in a nonlocal medium exhibits two distinct phases: a non-topological phase for small-amplitude solitons and quantized (topological) transport for solitons with moderate and large amplitudes. Additionally, we show that quantized transport can also occur for dipole and tripole solitons, but only within a finite range of propagation constants, beyond which the transport becomes irregular. In regions of quantized transport, soliton dynamics is governed by the space-time Chern numbers characterizing topological properties of the bands of the underlying linear system.

It should be mentioned that considerable interest to the formation of solitons in nonlocal nonlinear medium is mediated first of all by strong stabilizing action that nonlocality imparts not only on fundamental multidimensional solitons \cite{Turitsyn1986,Bang2002}, but also on higher-order excited self-sustained states, such as vortex solitons \cite{Briedis2005, Yakimenko2005, Skupin2006, Kartashov2007}, multipole and necklace solitons \cite{Kolchugina1980, Xu2005a, Rotschild2006, Kartashov2006}, and azimuthons \cite{Aguayo2006a, Aguayo2006b}. Such solitons have been widely studied and observed experimentally in materials with thermal nonlinearity \cite{Rotschild2005a, Rotschild2005b, Dreischuh2006}, where light absorption and heat transport processes are involved in the mechanism of nonlinear response, and in liquid crystals with reorientational nonlinearity \cite{Conti2003, Conti2004}. Nonlocality is also typical for photorefractive materials and for hot and cold atomic vapors. Importantly, some of these materials allow creation of periodic refractive index landscapes that can be electrically or optically controlled, see for example \cite{Fratalocchi2004, Fratalocchi2005, Assanto2007} and reviews \cite{Kartashov2009, Lederer2008}. Moreover nonlocality degree in such materials can be controlled too~\cite{Peccianti2005}, potentially leading to a plethora of interesting physical effects, such as enhancement of transverse mobility of solitons \cite{Xu2005b} in static lattices. {\color{black} For an overview of the physical significance of nonlocal media in optical and atomic systems, we refer the reader to recent reviews~\cite{Malomed2022, Mihalache2024}.}

\section{The model}
Propagation of a paraxial light beams in a nonlocal nonlinear medium with dynamically varying shallow refractive index can be described by the system of coupled dimensionless equations for light field amplitude $\psi$ and the nonlinear contribution to the refractive index $n$~\cite{Kartashov2009, Lederer2008, Peccianti2005, Xu2005b}  
\begin{eqnarray}
  	\label{NLS}
  	i\frac{\partial \psi}{\partial z}=-\frac{1}{2}\frac{\partial^2\psi}{\partial x^2} - \mathcal R(x,z)\psi -n\psi
  	\\
  	\label{n}
  	n-d\frac{\partial^2 n}{\partial x^2}=|\psi|^2.
\end{eqnarray}
Here $d$ is the degree of nonlocality of the nonlinear response ($d=0$  corresponds to local cubic medium, while $d\gg 1$ corresponds to strongly nonlocal regime); and the dynamical lattice is created by two periodic sublattices:  
\begin{eqnarray}
\label{R}
    \mathcal R(x,z)=p_1\cos^2\left(\frac{\pi x}{d_1}\right)+p_2\cos^2\left(\frac{\pi }{d_2}(x-\alpha z )\right)
\end{eqnarray}
with depths $p_{1,2}$, commensurate periods $d_{1,2}$, and sliding velocity $\alpha\ll 1 $ of the second sublattice with respect to the first one (its smallness is required for adiabaticity of the pumping process, when energy exchange between different bands of the lattice in the process of pumping does not occur). For commensurate $d_1$ and $d_2$, the lattice $\mathcal{R}(x,z)$ has a transverse period, which we denote by $X$ (for parameters selected above $X=d_2$), and longitudinal period $Z=d_2/\alpha$. \textcolor{black}{Such lattices can be created, e.g., in liquid crystals by applying a voltage to electrodes at the boundaries of the liquid crystal cell \cite{Fratalocchi2004, Fratalocchi2005, Assanto2007}. The created profile of the refractive index usually follows the geometry of electrodes. This also allows tuning of the nonlocality degree depending on the initial tilt of the liquid crystal molecules with respect to light propagation direction \cite{Peccianti2005}. While in previous experiments with lattice solitons such electrodes were straight, they can be made tilted and periodically joining to reproduce two mutually sliding patterns, periodically changing with distance $z$. When voltage is applied to such electrodes, this will result in $x$- and $z$-periodic reconfigurable distribution of the refractive index inside liquid crystal. } We also note that the system akin to (\ref{NLS}), (\ref{n}) can be used for the description of light propagation in lattices fabricated in semiconductors with nonlocal nonlinearities. \textcolor{black}{In this case, the arrays of waveguides are usually created by $x$-periodic indiffusion of dopants into material surface, and this process can also be adjusted to produce dynamical $z$-periodic indiffusion patterns creating sliding arrays.}

Equation (\ref{n}) can be solved with respect to $n$: 
\begin{eqnarray}
\label{nG}
  n(x,z)=\frac{1}{2\sqrt{d}}\int_{-\infty}^\infty e^{-|x-\xi|/\sqrt{d}}|\psi(\xi,z)|^2d\xi  
\end{eqnarray}
Thus, the nonlinear contribution to the refractive index in a given point depends on the intensity distribution in the entire transverse plane. In the limiting case $d\to 0$ one obtains the Kerr nonlinearity $n(x,z)=|\psi(x,z)|^2$. Further we will be interested in evolution of localized soliton solutions of the form $\psi=e^{ibz}w(x,z)$ and $n=n(x,z)$ in dynamically changing lattice $\mathcal{R}(x,z)$, where $w(x,z)$ and $n(x,z)$ are real squared integrable functions, and $|\partial_z w|, |\partial_z n|\ll b$. It follows from the direct integration of (\ref{n}) that the total power carried by any localized state described by (\ref{NLS}), (\ref{n}) is defined by 
\begin{eqnarray}
\label{power}
 U=\int_{-\infty}^{\infty} |\psi(x,z)|^2 dx =  \int_{-\infty}^{\infty} n(x,z)dx
\end{eqnarray}
and is $z-$independent. The location of a soliton in a lattice is determined by its center of mass $x_c(z)$:
\begin{eqnarray}
\label{xc}
    x_c(z)=\frac{1}{U}\int_{-\infty}^{\infty}x |\psi|^2dx=\frac{1}{U}\int_{-\infty}^{\infty}x ndx ,
\end{eqnarray}
(what is obtained by multiplying (\ref{n}) by $x$ and subsequent integration) while the the root mean-square widths of the soliton $L_\psi(z)$ and of the nonlinear contribution to the refractive index $L_n(z)$ are determined respectively by
$
    L_\psi^2=(1/U)\int_{-\infty}^{\infty}(x-x_c)^2 |\psi|^2dx,  
$
and 
$
    L_n^2=(1/U)\int_{-\infty}^{\infty}(x-x_c)^2 ndx.  
$
It follows directly from (\ref{n}) that
\begin{eqnarray}
L_n^2(z)-L_\psi^2(z)    =2d 
\end{eqnarray}
Thus, $L_n(z)>(2d)^{1/2}$ , i.e., for a given $z$ the width $L_n$ increases with the nonlocality degree at a rate of at least $d^{1/2}$.








\section{Linear spectrum}
\begin{figure}
\begin{center}
\includegraphics[width=\linewidth]{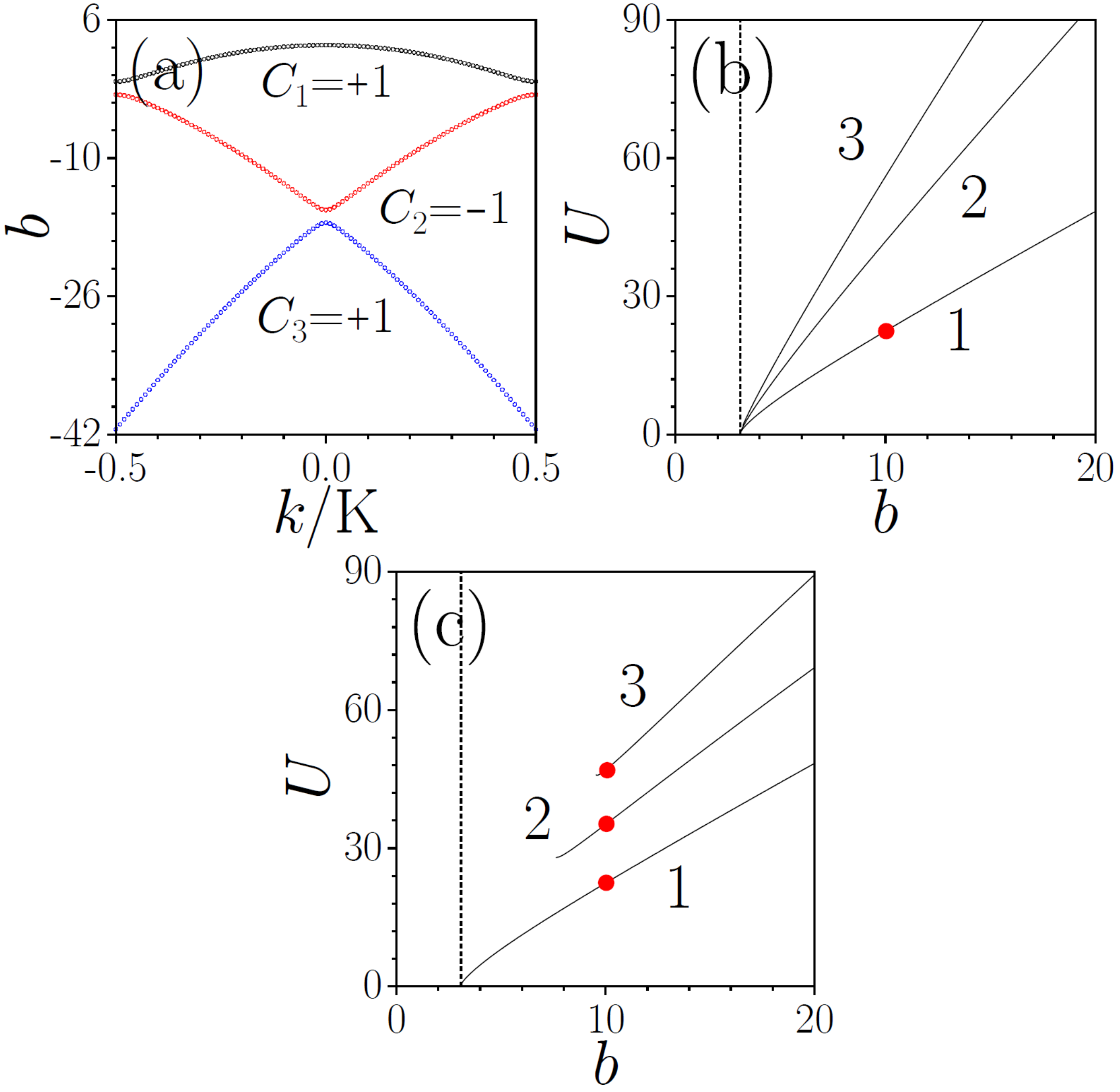}
 \end{center} 
  \caption{(a) Superimposed top three bands of the lattice calculated at different distances $z$ on one cycle $Z$ of lattice evolution. (b) $U(b)$ dependencies for fundamental lattice solitons calculated at $z=0$ for $d=1$ (curve 1), $d=5$  (curve 2), and $d=10$ (curve 3). (c) $U(b)$ dependencies for fundamental (curve 1), dipole (curve 2), and tripole (curve 3) lattice solitons for $d=1$. Red dots correspond to soliton profiles shown in Fig.~\ref{fig2}. Vertical dashed line indicates lower border of semi-infinite gap at $z=0$. Here and in all figures below $p_{1,2}=3$, $d_1=0.5$ and $d_2=1$.}
 \label{fig1}
\end{figure}

Since for static (or $z$-independent) lattices solitons are characterized by the propagation constants falling into one of the gaps of the underlying linear problem, it is instructive to consider the spectrum of the "instantaneous" Hamiltonian $H=-(1/2)\partial_x^2-\mathcal R$, i.e., $H\psi_{\nu k}=-b_{\nu k}\psi_{\nu k}$ with the lattice (\ref{R}) in which $z$ is considered as a parameter. Here $\psi_{\nu k}=u_{\nu k}\exp(ib_{\nu k}z+ikx)$ is the Bloch function, whose periodic part $u_{\nu k}(x,z)=u_{\nu k}(x+X,z)$ solves the equation
\begin{eqnarray}
\label{linear_eigen}
    b_{\nu k}u_{\nu k}=\frac{1}{2}\left(\frac{\partial^2 u_{\nu k}}{\partial x^2}+2ik\frac{\partial u_{\nu k}}{\partial x}-k^2u_{\nu k}\right)+\mathcal R u_{\nu k},
\end{eqnarray}
where $b_{\nu k}$ is the propagation constant, $k\in [-\rm K/2,\rm K/2)$ is the transverse Bloch momentum, $\textrm{K}=2\pi/X$ is the width of the first Brillouin zone, and $\nu$ is a band number (the upper band corresponds to $\nu=1$). The band structures obtained using plane-wave expansion methods at all distances $z$ within one pumping cycle, i.e., for $z\in[0,Z]$ are illustrated in Fig.~\ref{fig1}(a) for a particular case of lattice parameters $p_{1,2}=3$, $d_1=0.5$ and $d_2=1$, corresponding to the transverse lattice period $X=1$, and for sliding velocity $\alpha=0.01$ corresponding to $Z=100$.  

The bands do not cross for all $z$, what allows one to compute space-time Chern indices, characterizing topological properties of the bands, using the standard formula~\cite{Thouless1983}: 
 
\begin{eqnarray}
\label{chernindex}
C_\nu=\frac{1}{\pi}\textrm{Im}\int_0^Zdz\int_{{-\rm K}/2}^{{\rm K}/2}dk \langle\partial_k u_{\nu k}|\partial_z u_{\nu k}\rangle
\end{eqnarray}
where $\langle f|g\rangle=\int f^*gdx$. In Fig.~\ref{fig1}(a) we indicate the Chern indices for three upper bands. For our choice of parameters they are nonzero and have different signs for neighboring bands (although other choice of sublattice depths $p_{1,2}$ and periods $d_{1,2}$ may yield different magnitudes of the Chern indices).

\section{Solitons in a static lattice}
\begin{figure}
\begin{center}
\includegraphics[width=1\linewidth]{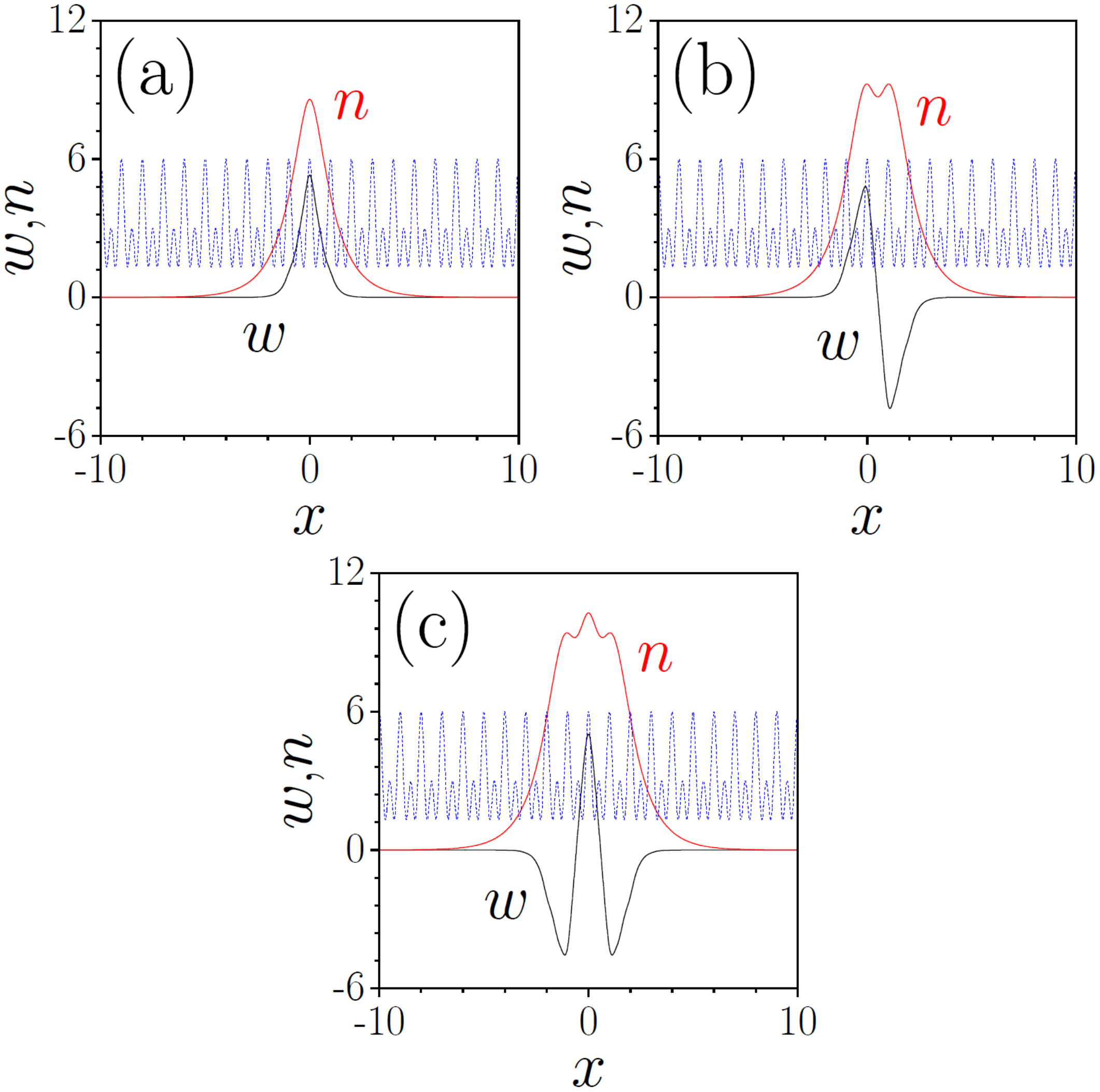}
\end{center} 
  \caption{Profiles of fundamental (a), dipole (b), and tripole (c) solitons with $b=10$  supported by the lattice at $z=0$ (or $\alpha=0$) and $d=1$. Red curves show refractive index distribution $n$, while blue dashed curves show the lattice profile. Solitons shown here correspond to the red dots in Fig.~\ref{fig1}.}
  \label{fig2}
\end{figure}

In a static lattice, i.e., at $\alpha=0$, solitons have the form $\psi=e^{ibz}w(x)$ and $n=n(x)$, where $w(x)$ and $n(x)$ do not depend on $z$. Such solitons can form in a semi-infinite gap $b>b_{1 k}$ or in one of finite gaps. We obtain them using the Newton iteration method for lattice profile $\mathcal R (x,z=0)$. Since in static lattices such solitons were studied in~\cite{Xu2005b}, here we discuss them only briefly. Besides the simplest single-hump (or fundamental) solitons illustrated in Fig.~\ref{fig2}(a), lattices with nonlocal nonlinearity support a number of higher-order excited states, such as dipole [Fig.~\ref{fig2}(b)] and tripole [Fig.~\ref{fig2}(c)] solitons, as well as more complex solutions with larger number of out-of-phase spots (solitons with in-phase spots are possible too, but they are unstable and we do not consider them here). In Fig.~\ref{fig2} we present also examples of the nonlinear contribution $n(x)$ to the refractive index for different types of soliton solutions. Soliton power $U$ defined by Eq. (\ref{power}) depends on the propagation constant $b$. Such dependence is shown in Fig.~\ref{fig1}(b) for fundamental solitons and in Fig.~\ref{fig1}(c) for dipole and tripole solitons. Fundamental solitons are thresholdless and bifurcate from the upper edge of the semi-infinite gap, i.e., from $b_{\nu=1,k=0}$ [shown by the vertical dashed line in Fig.~\ref{fig1}(b)].

Multipole solitons consist of out-of-phase spots with intensity maxima positioned near the neighboring strong maxima of the lattice (dashed blue lines in Fig.~\ref{fig2}), though they are slightly shifted relative to the lattice maxima. This shift increases as the propagation constant $b$ decreases. For a sufficiently large nonlocality degree, $d \gg 1$, a further decrease in $b$ can cause the soliton intensity maxima to shift into the neighboring strong maxima of  $\mathcal{R}$. This process continues as $b$ decreases, gradually transforming the multipole soliton into several well-separated, broad fundamental states that remain out of phase. Notably, this transformation occurs without any qualitative changes in the $U(b)$ dependence. In this case power $U$ monotonically decreases with decrease of $b$ and can even approach the edge of the gap, shown by the dashed line in Fig.~\ref{fig1}(c). Such transformations, however, do not occur for small and moderate nonlocality degree $d\sim 1 $ – in this case the families of multipole solitons do not reach the edge of the gap and exist only above certain power threshold [namely this situation is depicted in Fig.~\ref{fig1}(c)].

One can see from Fig. \ref{fig2} that despite strong intensity variations in multipole solitons, the refractive index distributions $n(x)$ remain smooth even at $d~\sim 1$ and can extend far beyond the region occupied by soliton due to nonlocality of the medium. As was shown above, for a fixed $b$ the width of $n(x)$ grows with $d$. Due to the broadening of the refractive index distribution with increasing $d$ at a fixed propagation constant $b$, the soliton profile expands too. This can be understood using (\ref{nG}). Indeed, assuming the opposite, i.e., that the width  $w(x)$ does not grow with $d$, for $d\gg L_\psi$, one could expand the exponential in (\ref{nG}) to obtain the estimate $n=Ud^{-1/2}+\mathcal{O}(d^{-1})$. Thus, in the leading order Eq.~(\ref{NLS}) would become linear and thus spatially localized solutions would be prohibited. This would contradict to the fact of existence of spatially localized nonlinear states. As a result, in our system the power $U$ increases with the degree of nonlocality as shown in Fig.~\ref{fig1}(b) (notice that in the absence of the lattice $U\propto \sqrt{d}$.)

\section{Thouless pumping of solitons}
We now turn to the dynamics of solitons induced by the sliding sublattices at $0<\alpha\ll 1$. Thus, as the initial conditions for Eq. (\ref{NLS}) we use exact soliton solutions obtained at $z=0$. We follow the $z-$evolution of the center of mass $x_c$ of solitons. Notably, the second equality in  (\ref{xc}) implies that the nonlinearly-induced refractive index moves with the same average velocity as the field $\psi$ that can be seen as a necessary condition for implementing the quantized transport. Another requirement for observation of quantized transport is the full population of one of the topological bands, which is not {\it a priori} guaranteed in optical systems, but depends on the initial conditions. Since all solitons considered here belong to the semi-infinite gap, they primarily populate the upper band with $\nu=1$. The degree of population of the band is determined by the soliton propagation constant, making the pumping dynamics highly sensitive to this parameter.

\begin{figure*}
\begin{center}
\includegraphics[width=\linewidth]{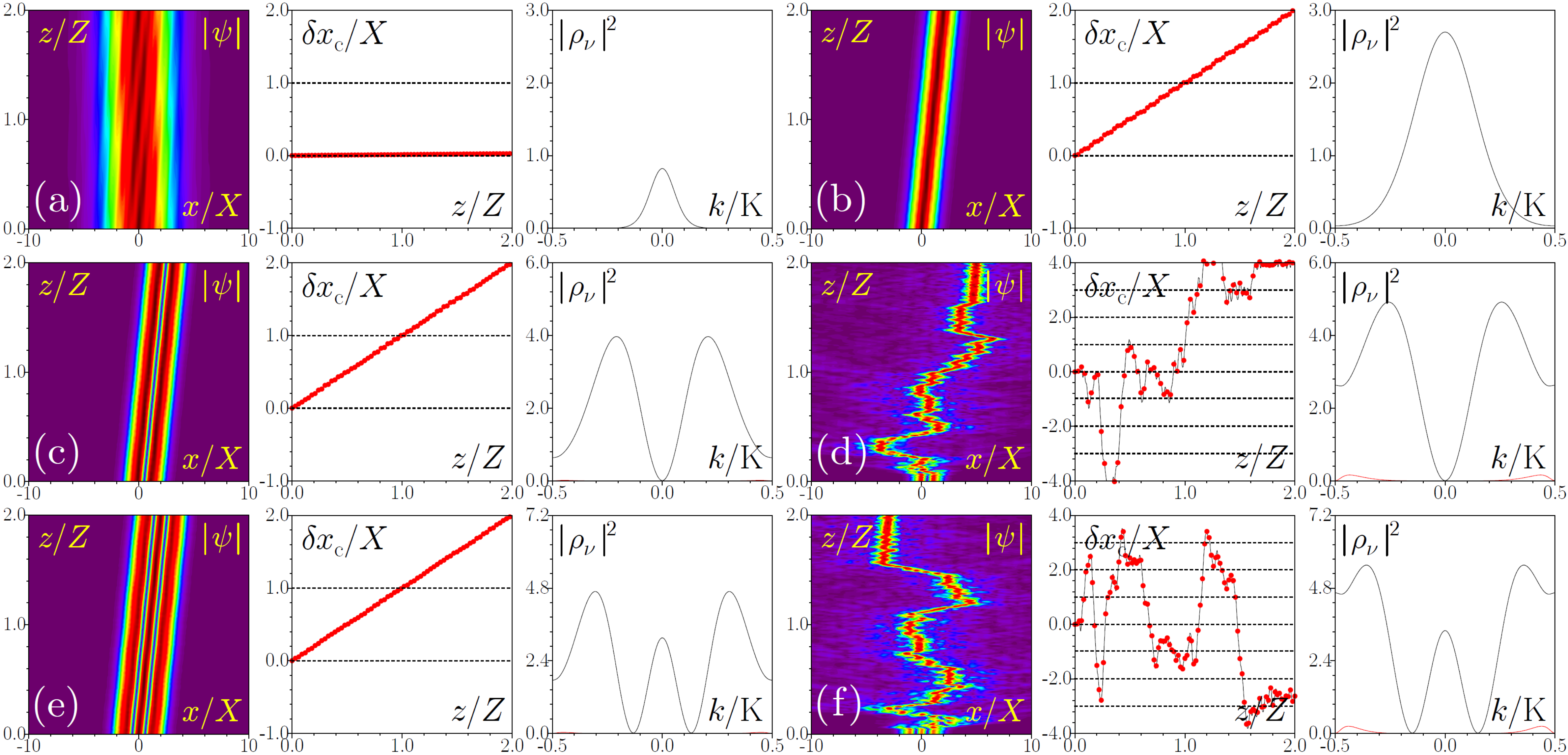}
\end{center}
\caption{Topological transport for fundamental (a),(b), dipole (c),(d), and tripole (e),(f) solitons in nonlocal medium at $d=5$, $\alpha=0.01$. Solitons correspond to $b=3.6$ (a), $9.2$  (b),  $14.0$ (c),  $21.0$ (d), $14.0$  (e), and $18.3$ (f). In each case, the field modulus distributions in the $(x,z)$ plane, soliton center displacement vs $z$, and projections on Bloch bands at $z=0$ are shown. In figures with projections black, red, and green lines show the projections on bands $\nu=1,2$, and $3$, respectively. The projection on band $\nu=3$ is not visible at the scale of the figure, while projection on band $\nu=2$ is always much smaller than that on band $\nu=1$. Panels (b), (c), (e) correspond to stable pumping.}
  \label{fig3}
\end{figure*}

Representative scenarios of the dynamics of propagation of fundamental solitons are 
presented in Fig.~\ref{fig3}(a) and~\ref{fig3}(b). In all cases, the first column shows propagation over two pumping cycles up to $z=2Z$, the second column shows center of mass displacement $\delta x_c=x_c(z)-x_c(0)$ versus distance $z$, while the third column shows projections $|\rho_\nu(k)|^2$ of the input field distribution, i.e., of the input soliton, on top three bands of the lattice within the first Brillouin zone (we show projections only at $z=0$ because they vary with $z$ only slightly due to the adiabaticity of the pumping process that signalizes on practical absence of power exchange between bands -- this is valid for all cases when soliton survives in the process of pumping). Notice that in second columns we show namely displacement $\delta x_c$ of soliton center of mass, because for example for dipoles $x_c(0)=X/2$, while for fundamental and tripole solitons $x_c(0)=0$. In Fig.~\ref{fig3}(a) we observe that a broad low-power soliton with a propagation constant close to the edge of the semi-infinite gap does not exhibit quantized transport, as the center of mass of such soliton experiences practically no displacement [see also the red line in the second panel of Fig.~\ref{fig3}(a)], even though the soliton experiences reshaping due to lattice variations. This is a consequence of incomplete population of the $\nu=1$ band, since such broad solitons populate only a narrow fraction of the band near Bloch momentum $k=0$ [see the right panel in Fig.~\ref{fig3}(a)]. In this situation the velocity of soliton is given by the group velocity of the Bloch mode $\psi_{1,k=0}$, which is zero. When propagation constant $b$ increases, the transition to stable pumping regime occurs, as illustrated in Fig.~\ref{fig3}(b) indicating that a narrow fundamental soliton populating practically the entire first band [see the right panel of Fig.~\ref{fig3}(b)] does experience topological transport; now the shift over one driving period $Z$ is given by the linear formula~\cite{Thouless1983} $\delta x_c=C_1Z$, where $C_1=1$ [see Fig. \ref{fig1}(a)].

\begin{figure}
\begin{center}

  \includegraphics[width=1.0\linewidth]{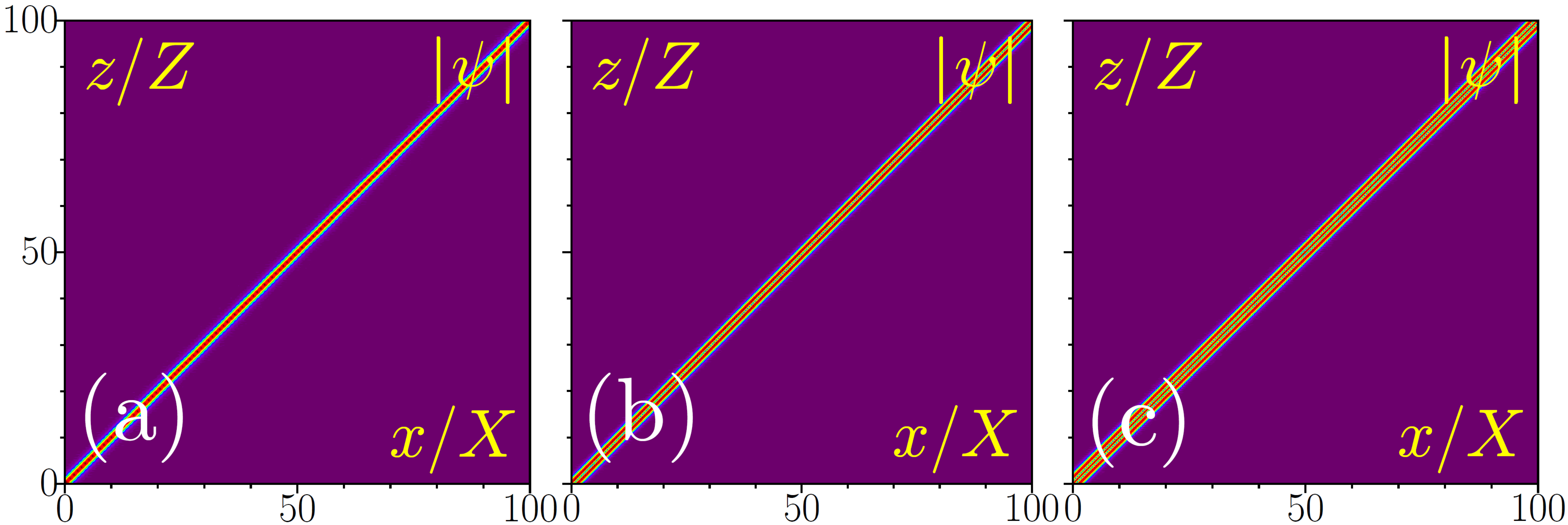}
\end{center}

\caption{Dynamics of stable topological transport over $100$ pumping cycles for fundamental soliton with $b=9.2$ (a), dipole soliton with $b=14.0$ (b), and tripole soliton with $b=14.0$ (c). In all cases $d=5$, $\alpha=0.01$.}
  \label{fig4}
\end{figure}

The described phase transition between non-topological and topological transport of solitons resembles a similar phenomenon that was recently reported for multi-frequency solitons~\cite{Kartashov2025}. We note that the region of non-topological transport near the band-edge, from which the soliton family bifurcates, becomes very narrow in sufficiently deep lattices, where in large diapason of amplitudes the solitons are well approximated~\cite{Alfimov2002} by the Wannier functions, while families are characterized by nearly linear $U(b)$ dependencies, like the one observed in Fig.~\ref{fig1}(b) and~\ref{fig1}(c). Evolution of such Wannier solitons is usually well described by discrete NLS equitation~\cite{Alfimov2002} that predicts quantized transport governed by the Chern number of the band from which the family bifurcates~\cite{Jurgensen2022}, but fails to describe the narrow region of non-topological behavior in the close vicinity of the band edge. The shallower is the lattice, the larger is the domain of nontopological pumping where solitons are described by the continuous NLS equation for the envelope of the carrier Bloch function. Notice also that Thouless pumping enabled by the nonlinearity was discussed in earlier studies~\cite{Mostaan2022,Kirsch2023,Viebahn2024}, where however the physical mechanism behind the effect of the nonlinearity was different from the one reported here.

\begin{figure}
\begin{center}
 \includegraphics[width=0.99\linewidth]{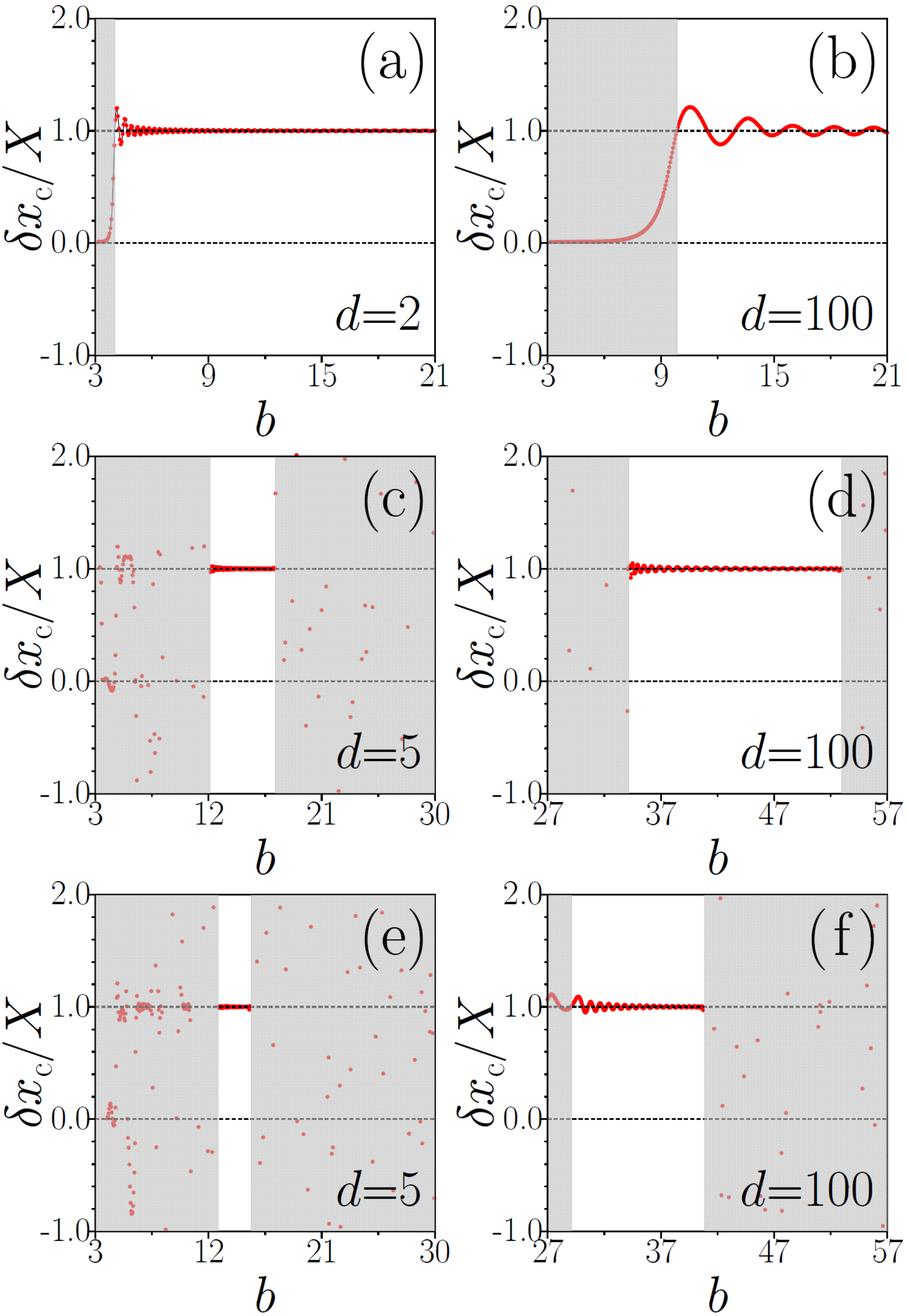}
\end{center} 
\caption{Center of mass displacement vs propagation constant (red dots) for fundamental solitons at $d=2$ (a), $d=100$ (b), for dipole solitons at $d=5$ (c), $d=100$ (d), and for tripole solitons at $d=5$ (e), $d=100$ (f). White regions show domains of stable pumping, while gray regions show domains where dynamics is not topological. In all cases $a=0.01$.}
  \label{fig5}
\end{figure} 

\begin{figure}
\begin{center}

  \includegraphics[width=0.99\linewidth]{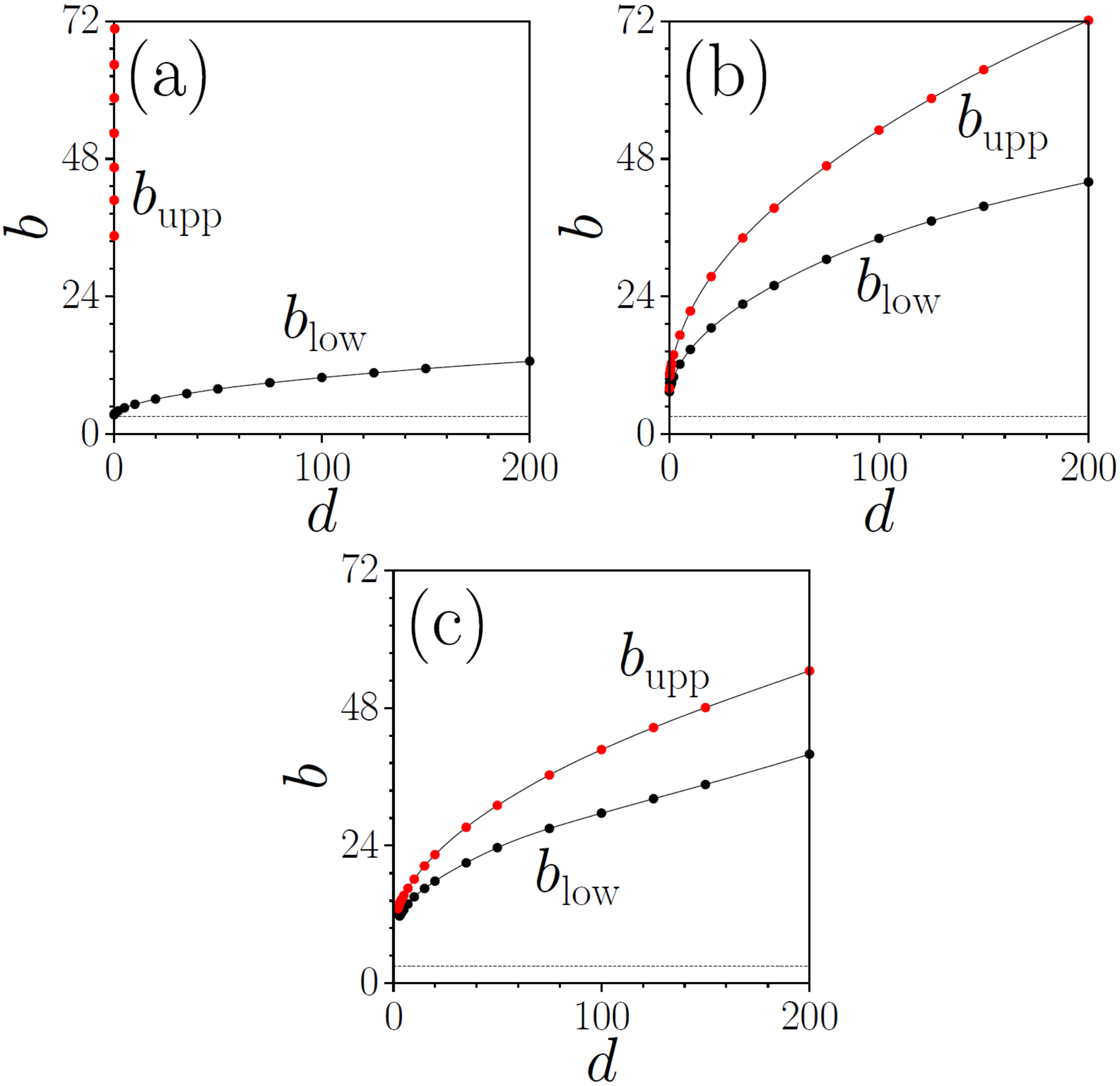}
\end{center}

\caption{Domains of stable topological pumping corresponding to $b_{\rm low}\leq b\leq b_{\rm upp}$ vs nonlocality degree $d$ for fundamental (a), dipole (b), and tripole (c) solitons. Horizontal dashed line indices lower border $b\approx 3.07$ of semi-infinite gap at $z=0$.}
  \label{fig6}
\end{figure}

The nonlocality of the nonlinear response significantly influences the aforementioned transition. Figures~\ref{fig5}(a) and \ref{fig5}(b) illustrate the dependence of the fundamental soliton center-of-mass displacement $\delta x_c(z)$ after one pumping cycle on the propagation constant $b$ in both moderate and strongly nonlocal regimes. The transition between regimes when no transport is observed (the gray region) and when stable pumping occurs (white region) is much sharper in medium with moderate nonlocality degree $d\sim 1$. Moreover, the increase of the nonlocality degree shifts the transition point between two regimes to larger propagation constants $b$. Hereafter we define the propagation constant $b=b_{\rm low}$, at which the displacement of the soliton center of mass at $z=Z$ reaches the value $\delta x_c=X$, as a lower border of stable pumping domain. The increase of $b_{\rm low}$ with $d$ can be explained as a result of the expansion of the nonlinear refractive index distribution, i.e., $L_n$, with increase of $d$ that leads to the expansion of the soliton profile discussed above, and, hence, narrowing of its projection on the upper spectral band despite the fact that the power of a soliton at a given $b$ increases with increase of the nonlocality degree. For the same reason, in Figs.\ref{fig5}(a) and~\ref{fig5}(b) at moderate and high $d$ values we do not observe breaking of pumping of high-power solitons, which is known to occur in local Kerr-type media~\cite{Jurgensen2021, Fu2022a, Fu2022b}. At $d=0$ breaking of pumping occurs at $b>b_{\rm upp}\approx 34.6$, when soliton starts oscillating around initial launching position and does not exhibit displacement. This value of $b_{\rm upp}$ can be interpreted as the upper border of stable pumping region. However, even slight increase of the nonlocality degree to $d=0.1$ shifts $b_{\rm upp}$ to the values exceeding 100 [see red dots in Fig. \ref{fig6}(a) showing $b_\textrm{upp}(d)$ dependence for fundamental solitons], effectively preventing breaking of pumping, in sharp contrast to the case of local medium.

One of our central results is that the nonlocality of the nonlinear response enhances the robustness of multipole solitons to such an extent that Thouless pumping can be observed in a nonlocal medium even for these higher-order excited states. The example of stable pumping of dipole soliton is presented in Fig.~\ref{fig3}(c). One can see that center of mass of such soliton exhibits quantized displacement, while soliton itself maintains its structural stability during pumping process despite dynamical variations of the lattice profile. We have found that such pumping can occur over hundreds of pumping cycles, i.e. that it is exceptionally robust process \textcolor{black}{(see examples of such pumping over $z=100Z$ cycles for fundamental, dipole and tripole solitons in Fig. \ref{fig4})}. Such solitons are characterized by wide, but structured projections $|\rho_\nu(k)|^2$ on the Bloch bands [Fig.~\ref{fig3}(c)]. \textcolor{black}{Notice that the number of peaks in projection on the band corresponds to the number of out-of-phase humps in soliton profile (i.e. there are two peaks for dipole solitons and three peaks for tripole solitons, see below). In addition to certain broadening of projection on the band with increase of propagation constant $b$ that reflects narrowing of individual poles in high-power solitons, we observe the increase of the amplitude of projection with $b$. It should be also stressed that the integral $\int_\textrm{BZ}|\rho_\nu(k)|^2 dk$ yielding population of the band $\nu$ has a very simple physical meaning - the sum of all such band populations is equal to soliton power $U$.} It turns out that stable pumping for dipole solitons occurs in limited interval of propagation constants. Figures~\ref{fig5}(c) and~\ref{fig5}(d) compare soliton center displacement vs $b$ dependencies in moderate and strongly nonlocal regimes. White regions in these plots correspond to the domain $b_{\rm low}\leq b\leq b_{\rm upp}$, where pumping is stable, while outside this domain the displacement of soliton center shows irregular behavior upon variation of $b$. This is because outside this domain dipole solitons exhibit dynamical instabilities that are accompanied by strong radiation and may even result in the formation of fundamental solitons that again show the tendency (after emission of radiation) to stable pumping [see Fig.~\ref{fig5}(d)]. Importantly, Fig.~\ref{fig5}(c) and~\ref{fig5}(d) illustrate notable expansion of stable pumping domain for dipole solitons with increase of the nonlocality degree of the medium.

A similar picture is observed for tripole solitons. The example of stable Thouless pumping for tripole soliton is presented in Fig.~\ref{fig3}(e). Such solitons have even more complex projection on the Bloch bands with three humps, but they still populate mainly the $\nu=1$ band. Inside stable pumping domain tripole solitons exhibit steady quantized displacement over hundreds of cycles \textcolor{black}{[see Fig. \ref{fig4}(c)]}, but outside it they exhibit instabilities usually leading to their decay and transformation into fundamental soliton, as shown in Fig.~\ref{fig3}(f). The resulting fundamental soliton start exhibit pumping [see later stages of propagation in Fig.~\ref{fig3}(f)]. The width of stable pumping domain for tripole solitons also increases with increase of nonlocality degree $d$ [compare Fig.~\ref{fig5}(e) and~\ref{fig5}(f)]. At the lower border of pumping domain in Fig.~\ref{fig5}(f) the poles of tripole soliton shift into next-neighboring maxima of the lattice leading to modification of the structure of solution, so we define $b_{\rm low}$ as a value at which soliton poles still remain in neighboring strong maxima of the lattice.

Stable pumping domains on the $(d,b)$ plane for all types of solitons considered here are compared in Fig.~\ref{fig6}. Dashed horizontal line in these figures denotes the border of the semi-infinite gap in lattice spectrum. One can conclude that fundamental solitons feature the broadest stable pumping domain [Fig.~\ref{fig6}(a)] that is in fact unlimited from above already for moderate nonlocality degree $d\sim 1$. The domain of stable pumping for dipole solitons is broader than that for tripole solitons [compare Fig.~\ref{fig6}(b) and~\ref{fig6}(c)]. Both these domains become wider with increase of the nonlocality degree. The width of stable pumping domain for dipole solitons shrinks to zero when $d\to 0$ while for tripole solitons stable pumping can be observed only starting from certain minimal nonlocality degree $d\approx 2.2$ [this is not visible at the scale of Fig.~\ref{fig6}(c)]. It should be mentioned that the decrease of the sliding velocity of two sublattices may affect the position of the lower border of stable pumping domain for fundamental solitons. Namely, with decrease of $\alpha$ the transition point $b_{\rm low}$ slightly shifts towards the edge of the gap (hinting on the fact that adiabaticity of pumping is essential for observed soliton dynamics). At the same time, decreasing the sliding velocity $\alpha$ {\em does not change} the borders of stable pumping domains for dipole and tripole solitons, i.e. the latter are determined by stability properties of such states in the dynamically varying lattice.

{\color{black}
\section{Some general considerations}

The model (\ref{NLS}), (\ref{n}) considered in this work, can be recast in the form
\begin{align}
	 \label{NLS_nonloc}
	 i\frac{\partial \psi}{\partial z}=&-\frac{1}{2}\frac{\partial^2\psi}{\partial x^2} - \mathcal{R}(x,z)\psi 
	 -\psi\int_{-\infty}^{\infty}K_d(x-y)|\psi(y,z)|^2dy 
\end{align}
where the exponential kernel is given by
\begin{align}
\label{kernel}
	K_d(x)=\frac{1}{2\sqrt{d}}e^{-|x|/\sqrt{d}} 
\end{align}
A general kernel function in (\ref{NLS_nonloc}), including (\ref{kernel}) satisfies the following normalization condition:
\begin{align}
\label{Knorm}
    \int_{-\infty}^\infty K_d(x)dx=1
\end{align}
This raises a general question on whether the above results on the adiabatic quantized transport can be applied to kernels, satisfying (\ref{Knorm}), of other types, say Gaussian or algebraic ones, provided Eq. (\ref{NLS_nonloc}) possess soliton solutions. While comprehensive answer to this question requires detailed numerical study of the respective model, for one-hump solitons some general conjectures can be formulated.

To this end we focus only on nonnegative kernels 
\begin{align}
    K_d(x)\geq 0
\end{align}
and search for solution of (\ref{NLS_nonloc}) in the form $\phi=e^{ibz}f(x,z)$. Using the adiabaticity of variation of potential, one can neglect $f_z$. This leads to the "instantaneous" (i.e., considered at a fixed  $z$) nonlinear problem
\begin{align}
	 \label{NLS_stat}
	 bf=&\frac{1}{2}\frac{\partial^2f}{\partial x^2} + \mathcal{R}(x,z)f
	 +f\int_{-\infty}^{\infty}K_d(x-y)|f(y,z)|^2dy 
\end{align}
The function $f$ satisfying (\ref{NLS_stat}) can be considered real.

Each such solution has a finite width $\ell(z)$ (see Appendix~\ref{sec:bound}). Thus, for a given total power $U$, for sufficiently small width of the kernel $d\ll \ell_{\rm min}$ where $\ell_{\rm min}=\min_{z\in Z} \ell (z)$,  one can explore the Taylor expansion 
\begin{align}
\label{Taylor}
    \int_{-\infty}^{\infty}K_d(x-y)|\psi(y,z)|^2dy \approx |\psi(x,z)|^2
     +K^{(1)}\frac{\partial|\psi(x,z)|^2}{\partial x}
     \nonumber \\ +K^{(2)}\frac{\partial^2|\psi(x,z)|^2}{\partial x^2}+\cdots
\end{align}
where
\begin{align}
    K^{(n)}=\frac{(-1)^n}{n!}\int_{-\infty}^\infty K_d(y)y^ndy
\end{align}
Thus, for any kernel $K_d(x)$ decaying at the infinity faster than any power of $x$, such that all $K^{(n)}$ exist, in the limit $d\to 0$ the leading order is reduced to conventional NLS equation (corrections to this equation can be accounted too). In this limit the differences between kernels (exponential, Gaussian, super-Gaussian, etc.) leads only to differences in the coefficients $K^{(n)}$, while the dynamics is expected to be qualitatively similar.

On the other hand, in the case of narrow kernels with algebraic decay, the condition $\lim_{d\to 0} K_d(x)\to\delta(x)$ (where $\delta(x)$ is the Dirac delta) implying also the required normalization (\ref{Knorm}), is not enough for conclusive comparison with the results obtained in this work, because (\ref{Taylor}) is not valid anymore.  

In the limit $d=\infty$ the spatially localized solutions of (\ref{NLS_stat}) do not exist. At large, but finite $d\to \infty$ Eq.~(\ref{Knorm}) implies the estimate $\max K_d(x)\sim 1/d\to 0$, which may hold also for algebraic kernels, like for example $K_d=d/[\pi(x^2+d^2)]$. Now for the existence of a solitonic solution with a given $U$, its width $\ell$ should grow with $d$ (otherwise the integral term in (\ref{NLS_stat}) would approach a constant value making Eq.~(\ref{NLS_stat}) linear and hence possessing no spatially localized solutions). Since $d$ and $\ell $ become now much larger than the period of the lattice $X$: $d, \ell\gg X $, it can be conjectured that the one-period shift becomes independent of the nonlocality parameter $d$ [this is what we observed in Figs.~\ref{fig5}(a) and (b) for the nonlocal media studied in this work]. These figures illustrate another common property of nonlocal kernels discussed in this section. Formation of a soliton with wide spatial spectrum, and hence stable quantized transport, require now larger $U$, and hence $b$, because $\max K_d(x)\sim 1/d\to 0$ at $d\to\infty$. These conjectures, however require further thorough numerical study, going beyond the scope of this work. 
}

\section{Conclusion}

In this paper we have extended the concept of topological soliton pumping to nonlocal nonlinear media. We have demonstrated that fundamental soliton transport exhibits two distinct phases: a non-topological phase at low power and a quantized phase at moderate and high power, where transport remains unbroken. Furthermore, we show that, within specific amplitude ranges, multipole solitonic states—such as dipole and tripole solitons—can also undergo pumping. Our findings are relevant not only to nonlinear optics, but also to atomic systems, where nonlocal interatomic interactions may play a crucial role and dynamically changing lattices can be easily created.  Our results demonstrate that even complex higher-order soliton states—such as dipoles and tripoles—can act as intrinsic topological carriers, with their stability and transport governed by the interplay of nonlocality and band geometry. Building on this, we expect that even more intricate topological states, such as optical vortex solitons, could potentially be pumped in nonlocal nonlinear two-dimensional materials. \textcolor{black}{The observation of the latter phenomenon would require creation in nonlocal medium, such as liquid crystal or material with thermal nonlinearity, of the dynamical $z$-periodic refractive index landscape consisting of two mutually sliding two-dimensional lattices with nontrivial topology of the bands, which, in addition should not overlap on one pumping cycle. In transparent dielectrics such dynamical two-dimensional landscapes can be inscribed using fs-laser writing. The potential alternative is infiltration of holes of $z$-varying photonic crystal fiber with liquids with nonlocal nonlinear response.}

\begin{acknowledgments}
Y.V.K. acknowledges funding by the research project FFUU-2024-0003 of the Institute of Spectroscopy of the Russian Academy of Sciences. V.V.K. was supported by the Portuguese Foundation for Science and Technology (FCT) under Contract No.2023.13176.PEX (DOI: https://doi.org/10.54499/2023.13176.PEX).
\end{acknowledgments}

\section*{Data Availability Statement}
The data that support the findings of this study are available from the corresponding author upon reasonable request.


%

\appendix 

{\color{black}
\section{On boundness of the soliton width}
\label{sec:bound}

Here we prove the finiteness of the soliton width $\ell$, expressed through the boundness of $\|f_x\|^2 \sim U/\ell^2$ ($\|\cdot\|$ denotes the conventional $L^2$-norm: $\|f_x\|^2:=\int_{-\infty}^\infty|f_x|^2dx$).
Using the known estimate: $\sup f^2\leq 2\|f\|\|f_x\|$ ($x\in\mathbb{R}$) valid for any real function $f$ (we assume that $f$ is differentiable and all integrals used here and below exist) as well as normalization (\ref{Knorm}), multiplying (\ref{NLS_stat}) by $f$ and integrating we obtain the set of relations
\begin{align}
    \frac{1}{2}\|f_x\|^2=&-b\|f\|^2+ \int_{-\infty}^{\infty}dx \mathcal{R}(x,z)f^2 (x,z)
	\nonumber\\ 
	&+\iint_{-\infty}^{\infty}dxdyK_d(x-y)|f(x,z)|^2|f(y,z)|^2
    \nonumber \\
    &\leq -b\|f\|^2+\mathcal{R}_m \|f\|^2+\|f\|^2\sup f^2
    \nonumber \\
    &\leq (\mathcal{R}_m  -b) \|f\|^2+2\|f\|^3 \|f_x\|
     \nonumber \\
    &\leq (\mathcal{R}_m  -b)\|f\|^2+\eta \|f\|^6+\frac{\|f_x\|^2 }{\eta} 
\end{align}
Here $\mathcal{R}_m=\max\{\mathcal{R}(x,z)\}$ and we used that
\begin{align}
    2\|f\|^3\|f_x\|\leq \eta \|f\|^6+ \frac{ \|f_x\|^2}{\eta}  
\end{align}
where $\eta>0$ can be any positive number.  Rearranging the terms we obtain
\begin{align}
    \left(\frac{1}{2}-\frac{1}{\eta}\right) \|f_x\|^2 \leq (\mathcal{R}_m  -b)U+\eta U^3 
\end{align}
(recall that $\|f\|^2=U$). Hence, by considering $\eta>2$ one obtains an upper estimate for the "kinetic energy" $\|f_x\|^2$, and consequently the lower bound of the soliton width $\ell$ for given $z$ and $U$ (the propagation constant is a function of $U$).

}

\end{document}